\documentstyle[prd,aps,epsf]{revtex}
\begin{document}
\makeatletter
 \newdimen\ex@
 \ex@.2326ex
 \def\dddot#1{{\mathop{#1}\limits^{\vbox to-1.4\ex@{\kern-\tw@\ex@
  \hbox{\tenrm...}\vss}}}}
 \makeatother

\thispagestyle{empty}
{\baselineskip0pt
\leftline{\baselineskip16pt\sl\vbox to0pt{\hbox
{\it Department of Physics}
               \hbox{\it Waseda University}\vss}}
\rightline{\baselineskip16pt\rm\vbox to20pt{\hbox{WU-AP/136/01}
%           \hbox{December 1998}
%            \hbox{\today} 
\vss}}%
}
\vskip1cm
%\vfill
\begin{center}{\large \bf
Reconstructing the equation of state for cold nuclear matter
from the relationship of any two properties of neutron stars
}
\end{center}
\vskip1cm
\begin{center}
 {\large 
Tomohiro Harada
\footnote{Electronic address: harada@gravity.phys.waseda.ac.jp}\\
{\em Department of Physics,~Waseda University, Shinjuku, 
Tokyo 169-8555, Japan}}
\end{center}
%\vfill

\begin{abstract}
A direct method is developed to 
reconstruct the equation of state for
high-density nuclear matter from the relationship 
between any two properties of
neutron stars, such as
masses, radii, moments of inertia, baryonic masses, binding energies,
gravitational redshifts, and their combinations.
\end{abstract}


The equation of state (EOS) for 
supranuclear-density matter with zero temperature
is relevant to neutron stars.
For supranuclear-density matter,
a number of possible scenarios, such as
Kaon condensation, pion condensation,
hyperonic matter, and strange quark matter,
have been proposed.
In the near future, lattice QCD is expected to 
predict the correct EOS for supranuclear-density matter.
However, an accelerator-experimental test of the theoretical 
predictions for the EOS for cold high-density 
nuclear matter is extremely difficult.

In this light, neutron stars are very unique 
laboratories.
Neutron stars are observed as pulsars, X-ray compact sources,
soft $\gamma$-ray repeaters, and gravitational wave sources.
Binary pulsar observations can 
determine masses of neutron stars with very high 
accuracy~\cite{thorsett1999}.
From observations of type-I X-ray bursts, kHz 
quasiperiodic oscillations (QPO's) of 
low mass X-ray binaries,
and so on, some constraints on 
the radius of neutron stars have been obtained.
Actually, the value of the radius itself 
has been already measured~\cite{rutledge2000}.
Very recently, it has been proposed that 
the radius of neutron stars
can be directly measured by
the observation of gravitational waves
from coalescing neutron star (NS) 
/black hole (BH) binaries~\cite{vallisneri1999,saijo2000}.
Some combinations of masses and spins of coalescing NS/NS
and NS/BH binaries can be determined from 
gravitational wave observations
in the inspiralling phase~\cite{cutler1996}.
Recently, it has been pointed out 
that the broad peaks at frequencies $\sim 20-40~\mbox{Hz}$ 
of kHz QPO's can be interpreted as
the precession frequency due to the Lense-Thirring effect,
the relativistic frame-dragging effect due to
the spin angular momentum~\cite{stella1999}.
This suggests the possibility of direct detection of 
the ratio of the moment of inertia to the mass 
of neutron stars. 
Observations of absorption lines in X-ray spectra
may reveal the gravitational 
redshift~\cite{lattimer1990,glendenning1992}.
In some situations,
from knowledge of progenitors of supernovae,
we may infer the baryonic mass of 
the remnant neutron stars.
In next decade, we will have a good amount of data on
the properties of neutron stars.

The structure of neutron stars has been investigated
for a large set of EOS's.
Arnett and Bowers~\cite{arnett1977} examined systematically
the mass-radius relation of neutron stars for various 
EOS's.
Recently, Lattimer and Prakash~\cite{lattimer2000}
examined systematically the relation of masses, radii, moments of
inertia, and binding energies for a large number of 
modern EOS's.
An alternative approach is a direct reconstruction
of the EOS from the observed properties of
neutron stars.
Lindblom~\cite{lindblom1992} developed a method to 
reconstruct the EOS for nuclear matter
from neutron-star masses and radii.
In this report, we give a
direct method to reconstruct the EOS
from the relationship of any two properties of 
neutron stars.
We adopt geometrized units such that $c=G=1$.

Before the method of reconstruction is derived, it is helpful
to briefly review how observable quantities of neutron stars
are obtained for a given EOS.
For simplicity, we assume that neutron stars 
are static and spherically symmetric with no magnetic field.
The line element is given by
\begin{equation}
ds^{2}=-e^{\nu(r)}dt^{2}+e^{\lambda(r)}dr^{2}
+r^{2}(d\theta^{2}+\sin^{2}
\theta d\phi^{2}).
\end{equation}
The structure of neutron stars is determined by the 
Tolman-Oppenheimer-Volkoff (TOV) equations given by
\begin{eqnarray}
P'&=&-(P+\epsilon)\frac{m+4\pi r^{3}P}{r(r-2m)},\\
m'&=&4\pi \epsilon r^{2} \\
\nu'&=&-\frac{2}{P+\epsilon}P',
\end{eqnarray}
where $P$ and $\epsilon$ are the
pressure and energy density 
of the matter, $m$ is defined as
\begin{equation}
1-\frac{2m}{r}=e^{-\lambda},
\end{equation}
and the prime denotes the derivative with respect to $r$.
For a given EOS, it is easy to integrate the TOV equations
with the initial values $P(0)=P_{c}$ and $m(0)=0$.
The radius $R$ is the value of $r$ such that $P=0$ at $r=R$.
The mass $M$ is obtained by $M=m(R)$.
The moment of inertia $I$ is calculated as~\cite{hartle1967}
\begin{equation}
I=\frac{1}{6}R^{4}\frac{\omega'(R)}{\omega(R)},
\end{equation}
where $\omega(r)$ is obtained by integrating the following
ordinary differential equation 
\begin{equation}
\frac{1}{r^{4}}(r^{4}e^{-\frac{\lambda+\nu}{2}}\omega'
)'+\frac{4}{r}(e^{- \frac{\lambda+\nu}{2}})'
\omega=0,
\end{equation}
with the initial value $\omega(0)\ne 0$ and $\omega'(0)=0$.
The baryonic mass $m_{B} N$ and binding energy $B$ are obtained by
\begin{eqnarray}
m_{B}N&=& \int^{R}_{0}dr e^{\frac{\lambda}{2}}4\pi\rho r^{2}, \\
B&=&m_{B}N-M,
\end{eqnarray}
where $\rho$ is the rest-mass density that is written as
$\rho=m_{B}n$ using the baryon mass $m_{B}$ and the baryon number 
density $n$. 
The gravitational redshift $z$ at the surface is given by
\begin{equation}
z=\left(1-\frac{2M}{R}\right)^{-1/2}-1
\end{equation}

Then, we describe how to reconstruct the EOS from a complete
observed data set.
The problem is to determine the EOS $\rho=\rho(P)$
for the domain $P_{0}\le P \le P_{U}$, where
$P_{U}$ is some cutoff pressure.
We assume that we have a complete relationship 
$f(Q^{(1)}, Q^{(2)})=0$ between 
observables $Q^{(1)}$ and $Q^{(2)}$
and that we have the correct EOS 
for $P\le P_{0}$.
We also assume that $\rho$ is a nondecreasing function of $P$
\footnote{Otherwise, the fluid is locally unstable.}.
We prepare $N$ grid points in the domain $[P_{0},P_{N}]$
as $P_{0}<P_{1}<\cdots<P_{i}<P_{i+1}<\cdots<
P_{N}=P_{U}$.
Here, as a simple example, we take the grid points as
\begin{equation}
\log P_{i}=\log P_{0}+\frac{i}{N}
(\log P_{N}-\log P_{0}).
\end{equation}

Suppose we have the ``pointwise exact'' 
EOS for $P\le P_{i}$.
This implies that if the TOV equations are integrated
with the initial values $P(0)=P_{k}$ and $m(0)=0$
then
the obtained $Q^{(1)}_{k}$ and $Q^{(2)}_{k}$ 
satisfy $f(Q^{(1)}_{k},Q^{(2)}_{k})=0$
for $0\le k\le i$.
Using some unknown number $\rho_{i+1}=\rho(P_{i+1})$, we interpolate
the EOS in the interval $[P_{i},P_{i+1}]$.
If it is taken into account that the EOS for nuclear matter
is locally close to polytropic, then,
the best interpolation is the logarithmically 
linear interpolation
\begin{equation}
\log P=\log P_{i} +\Gamma_{i}(\log \rho -\log \rho_{i}),
\end{equation}
where $\Gamma_{i}$ is the ``local polytropic index'' given by
\begin{equation}
\Gamma_{i}\equiv
\frac{\log P_{i+1}-\log P_{i}}{\log \rho_{i+1}-\log \rho_{i}}.
\label{eq:gammai+1}
\end{equation}
Using this interpolation, $\epsilon$ is obtained as
functions of $P\in [P_{i},P_{i+1}]$ as
\begin{equation}
\epsilon=\rho(1+e), 
\end{equation}
where 
\begin{eqnarray}
\rho&=&\left(\frac{P}{K_{i}}\right)^{\Gamma_{i}^{-1}}, \\
e&=&\frac{K_{i}}{\Gamma_{i}-1}\left[\left(\frac{P}{K_{i}}\right)
^{1-\Gamma_{i}^{-1}}-\rho_{i}^{\Gamma_{i}-1}\right]+e_{i}, 
\label{eq:e}\\
K_{i}&\equiv &\frac{P_{i}}{\rho_{i}^{\Gamma_{i}}},
\label{eq:Ki}
\end{eqnarray}
$e$ is the specific internal energy and 
Eq.~(\ref{eq:e}) is obtained from the first law of thermodynamics.
Now that we have the EOS for $P\le P_{i+1}$
parametrized by one parameter $\rho_{i+1}$, we construct
the neutron star
by integrating the TOV equations with the initial values 
$P(0)=P_{i+1}$ and $m(0)=0$.
Since the obtained $Q^{(1)}_{i+1}$ and $Q^{(2)}_{i+1}$ are 
functions of $\rho_{i+1}$, we denote them as
$Q^{(1)}_{i+1}(\rho_{i+1})$ and $Q^{(2)}_{i+1}(\rho_{i+1})$.
It is easy to numerically find 
the root $\rho_{i+1}$ ($\ge \rho_{i}$) 
which gives $ f(Q^{(1)}_{i+1}(\rho_{i+1}),Q^{(2)}_{i+1}(\rho_{i+1}))=0$,
by the bisection method or
Newton-Raphson method
with the aid of numerically estimated derivative
$d f(Q^{(1)}_{i+1}(\rho_{i+1}),
Q^{(2)}_{i+1}(\rho_{i+1}))/d \rho_{i+1}$.
It is noted that $(\rho_{i+1}-\rho_{i})$ may not always be
small even for small $(P_{i+1}-P_{i})$
because of the possible first order phase transitions. 
Thus, we find $e_{i+1}$ as
\begin{equation}
e_{i+1}=\frac{K_{i}}{\Gamma_{i}-1}
(\rho_{i+1}^{\Gamma_{i}-1}-\rho_{i}^{\Gamma_{i}-1})+e_{i}.
\end{equation}
Then we obtain the pointwise exact EOS 
for $P\le P_{i+1}$.

Repeating these processes, we reconstruct the 
EOS for $P\le P_{N}=P_{U}$.
The reconstructed EOS is pointwise exact 
in the sense that the relationship $f(Q^{(1)},Q^{(2)})=0$
is exactly satisfied by
all neutron stars with the central pressure
$P\in\{P_{0}, P_{1}, \cdots\,P_{N}\}$.
By taking the limit $N\to\infty$, 
the reconstructed EOS becomes exact
and in this limit the reconstruction does not depend on 
the interpolation method. 

In practice, it is very important how fast this scheme converges 
as $N$ is increased. It is clear that each $i$th step
reproduces the correct answer if the EOS in the
interval $[P_{i},P_{i+1}]$ is a polytropic one.
This implies that our extrapolation is linear in each interval
with respect to $\log P$ and $\log \rho$. For an estimate,
we assume that $\log \rho$ is Taylor expandable with respect to 
$\log P$ as
\begin{equation}
\log \rho = \log \rho_{i}+\left(\frac{d\log \rho}{d\log P}\right)_{i}
(\log P-\log P_{i})
+\frac{1}{2} \left(\frac{d^{2}\log \rho}{(d\log P)^{2}}\right)_{i}
(\log P-\log P_{i})^{2}
+\cdots.
\end{equation}
The error produced in the interval $[P_{i},P_{i+1}]$ is estimated 
for sufficiently large $N$ as
\begin{equation}
(error)_{i}\simeq
\left(\frac{d^{2}\log \rho}{(d\log P)^{2}}\right)_{i}
(\log P_{i+1}-\log P_{i})^{2}.
\end{equation}
Then the totally accumulated error is bounded from above as
\begin{equation}
\left|\sum_{i=0}^{N-1}(error)_{i}\right|
\alt \frac{1}{N}(\log P_{N}-\log P_{0})\sup_{[P_{0},P_{N}]}
\left|\frac{d^{2}\log \rho}{(d\log P)^{2}}\right|.
\end{equation}
Therefore the convergence is as fast as $1/N$. 

Next, we describe the reconstruction from a finite data set.
In this case, we have only 
a set of $N$ reliable data pairs
\begin{equation}
\{(Q^{(1)}_{1},Q^{(2)}_{1}),\cdots,
(Q^{(1)}_{i},Q^{(2)}_{i}),\cdots, (Q^{(1)}_{N},Q^{(2)}_{N})\}.
\end{equation} 
The order of data pairs should be determined so that 
the smaller value of index $i$ corresponds to 
a neutron star with the lower central pressure.
The order of data pairs does not seem to be determined {\it a priori}
except for the fact that 
$M$ and $m_{B}N$ are increasing functions of the central pressure
for stable neutron stars.
Unlike the previous case, we have to determine two unknown
numbers $P_{i+1}$ and $\rho_{i+1}$ such that
\begin{eqnarray}
Q^{(1)}_{i+1}(P_{i+1},\rho_{i+1})&=&Q^{(1)}_{i+1}, \\
Q^{(2)}_{i+1}(P_{i+1},\rho_{i+1})&=&Q^{(2)}_{i+1},
\end{eqnarray}
where $Q^{(1)}_{i+1}(P_{i+1},\rho_{i+1})$ and 
$Q^{(2)}_{i+1}(P_{i+1},\rho_{i+1})$ are the values of $Q^{(1)}$
and $Q^{(2)}$, respectively,
which are obtained by integrating the TOV equations
with $P(0)=P_{i+1}$ and $m(0)=0$ with the EOS
interpolated by Eq.~(\ref{eq:gammai+1})-(\ref{eq:Ki})
using the unknown numbers $P_{i+1}$ and $\rho_{i+1}$.
This can be done numerically by the two-dimensional 
Newton-Raphson method 
with the aid of the numerically evaluated Jacobian 
\begin{equation}
\frac{\partial(Q^{(1)}_{i+1}(P_{i+1},\rho_{i+1}),
Q^{(2)}_{i+1}(P_{i+1},\rho_{i+1}))}{\partial(P_{i+1},\rho_{i+1})}.
\end{equation} 
Then we obtain the pointwise exact EOS 
for $P\le P_{i+1}$.
If a set of $\log P_{i}$ has roughly homogeneous 
distribution, the convergence 
is as fast as $1/N$
as we have already seen. 

Here, we briefly discuss desired observational accuracy
in order to draw some useful conclusions
in discriminating between the possible EOS's,
based on the figures given by Lattimer and Prakash~\cite{lattimer2000}. 
The data pairs on masses and 
radii of neutron stars will be very powerful 
if the measurement is as accurate as
$\Delta M \alt 0.1 M_{\odot}$ and $\Delta R\alt 1~\mbox{km}$. 
For the pairs $I/MR^{2}$ and $M/R$, 
$\Delta (I/MR^{2}) \alt 0.01$ and $\Delta (M/R) \alt 0.01$ 
would be desired.
For the pairs $B/M$ and $M/R$, 
$\Delta (B/M) \alt 0.01$ and $\Delta (M/R) \alt 0.01$ 
would be desired.
However, the above estimate is quite naive, and 
more systematic and elaborate study is needed.
The present accuracy for mass measurement from a radio binary pulsar
is $\Delta M \simeq 0.0007 M_{\odot}$~\cite{thorsett1999},
while that from an X-ray binary is 
$\Delta M \simeq 0.4 M_{\odot}$~\cite{orosz1999}.
For the measurement of radius, 
$\Delta R\simeq 5.2~\mbox{km}$ is obtained if
the interstellar absorption of X-ray emission and
the distance to the object are fixed~\cite{rutledge2000}.
More detailed X-ray spectral analyses and/or the space interferometric 
mission
will dramatically improve the accuracy in determining the 
radii of neutron stars.
Although moments of inertia of neutron stars have not been 
determined directly,
the determination of them 
will be a complementary and potentially powerful technique.

Finally, we compare the present method of reconstruction
from a complete data set
with that of Lindblom~\cite{lindblom1992}'s one.
(1) The former applies to any two properties of neutron stars
while the application of the latter is restricted to 
the mass-radius relation. 
(2) The former needs iteration to determine $\rho_{i+1}$
for each $i$th step,
while $P_{i+1}$ and $\epsilon_{i+1}$ are determined
explicitly and no iteration is needed in the latter.
(3) In the former, 
the relation $f(Q^{(1)},Q^{(2)})=0$ is satisfied at any grid point 
$P\in\{P_{0},P_{1},\cdots, P_{N}\}$ for finite $N$,		
while it is not in the latter.
(4) In the former, the obtained EOS 
is of the form $P=P(\rho)$, which is conventional
in nuclear physics, while in the latter it is of the form
$P=P(\epsilon)$.

In summary, a complete data set of
two properties of neutron stars is sufficient to
reconstruct the EOS for cold nuclear matter.
We have proposed a direct method 
to reconstruct the EOS.
It is crucial to 
measure more than one property of a neutron star
in order to put stringent constraints on
the EOS.

I would like to thank
T.~Tatsumi, K.~Nakao, M.~Yasuhira, K.~Ioka, and H.~Sotani
for helpful discussions and comments.
I am very grateful to J.~Overduin for carefully reading the 
manuscript.
I am also grateful to K.~Maeda for continuous encouragement.
This work was supported by the 
Grant-in-Aid for Scientific Research (No. 05540)
from the Japanese Ministry of
Education, Culture, Sports, Science, and Technology.

\end{document}